\documentstyle[11pt,psfig]{article}
\begin{document}
\baselineskip=\normalbaselineskip\multiply\baselineskip
by 150\divide\baselineskip by 100
\pagenumbering{arabic}
\pagestyle{plain}
\reversemarginpar
\begin{titlepage}
\begin{flushright}
ANL--HEP--PR--95--66\\
MSUHEP--50925
\end{flushright}
\begin{flushright}
September 1995
\end{flushright}
\vspace{0.4cm}
\begin{center}
\large
{\bf Detecting a Light Stop from Top Decays \\
at the Tevatron}
\end{center}
\begin{center}
{\bf S. Mrenna}\footnote{mrenna@hep.anl.gov}
\end{center}
\begin{center}
{High Energy Physics Division\\
Argonne National Laboratory \\
Argonne, IL  60439}
\end{center}
\begin{center}
and
\end{center}
\begin{center}
{\bf C.--P. Yuan}\footnote{yuan@msupa.pa.msu.edu}
\end{center}
\begin{center}
{Department of Physics and Astronomy \\
Michigan State University \\
East Lansing, MI 48824}
\end{center}
\vspace{0.4cm}
\raggedbottom
\setcounter{page}{1}
\relax

\begin{abstract}
\noindent
We study the possibility of discovering or excluding
a light top squark (stop) $\tilde{t}_1$
based on top quark decays in the $t\overline{t}$
events produced at the Fermilab Tevatron.
In particular, we consider the Minimal Supersymmetric
Standard Model with the sparticle
spectrum  $m_{\chi^\pm_1}+m_b, \, M_W+m_{\chi^0_1}+m_b
> m_{\tilde{t}_1} >m_{\chi^0_1}+m_c$,  where
$\chi^0_1$ is the lightest neutralino, so that
$t\rightarrow\tilde{t}_1 \chi^0_1$ and
$\tilde{t}_1\rightarrow c\chi^0_1$.  All other
sparticle masses are assumed to be heavier than $m_t$.
Such a spectrum seeks to explain
the experimental values of $\alpha_s(M_Z^2)$, $R_b$ and $A_{LR}$
obtained from LEP/SLC data.
We find that the prospect to observe a light stop via this channel
at the Tevatron is very promising.\\ \\
PAC codes:  12.15.Mm, 14.65.Ha, 14.80.L
\end{abstract}

\end{titlepage}
\newpage

\section{Introduction}
\indent

A recent analysis of the strong and electroweak parameters extracted from
LEP/SLC data has shown that there is better agreement between
theory and experiment for the Minimal Supersymmetric Standard Model (MSSM)
with light superpartners than for
the Standard Model (SM) alone \cite{stuart}.
If the superpartners of the MSSM are light, then new top decay modes can
exist.   There are many compelling reasons to believe that the lightest
superpartner is the lightest neutralino $\chi^0_1$ (see reference
\cite{report} for a review.)   A reconciliation of the strong coupling
measurement at the scale ${M}_{Z}$, $\alpha_s(M_Z^2)$, the observed
branching ratio of
${Z}^0\rightarrow b\overline{b}$, $R_b$,
and the left-right asymmetry measured at the SLC, $A_{LR}$,
is suggested by the MSSM where
the lightest top squark
(stop) $\tilde{t}_1$, the lightest chargino $\chi^\pm_1$, and
the lightest neutralino $\chi^0_1$ have masses below $m_t$,
while all other superpartners are relatively heavy.
As shown in Fig.~2 of Ref.~\cite{stuart},
a large $m_{\chi^\pm_1}$ would preferentially require the
branching ratio (BR)
of $t \rightarrow {\tilde{t}_1} {\chi^0_1}$ to be small.
(The small branching ratio of any non--SM decay mode
of the top quark is inferred from the current
top quark data at the Tevatron. This will be shown in Sec. 3.)
Over a large range of this parameter space,
$m_{\chi^\pm_1} > m_{\tilde{t}_1}$ (see Fig. 1 of Ref.~\cite{stuart}).
Therefore, the scenario arises where
$t\rightarrow \tilde{t}_1 \chi^0_1$ provided that $m_t > m_{\tilde{t}_1}+
m_{\chi^0_1}$.  Furthermore, if $m_{\tilde{t}_1} < m_{\chi^\pm_1}+m_b$,
and  $M_W+m_{\chi^0_1}+m_b > m_{\tilde{t}_1} >m_{\chi^0_1}+m_c$,
then the dominant decay of $\tilde{t}_1$ is
$\tilde{t}_1\rightarrow c \chi^0_1$~\cite{bdggt}. This is
a flavor--changing--neutral--current decay.

Also recently, an analysis has been performed
by the {D\O} collaboration to
exclude light stop squarks from the jet and missing transverse energy
signature expected from stop squark pair production \cite{claes}.
The D\O~analysis relies on the large cross section for pair production of
light stop squarks via the partonic processes $q\overline{q},gg\rightarrow
\tilde{t}_1\tilde{t}^*_1$ and the distinctive final state
$\tilde{t}_1\tilde{t}^*_1\rightarrow c\overline{c}\chi^0_1\chi^0_1$ when
$m_{\chi^\pm_1} + m_b > m_{\tilde{t}_1}$.
The signature is two acollinear jets and large missing transverse energy
(${\hbox to -2pt{/\hss}}E_T$).  This signal is overshadowed by the
background, however,
when the $c$--jet and/or
${\hbox to -2pt{/\hss}}E_T$ spectra becomes soft.  As a result,
the D\O~analysis excludes the $(m_{\tilde{t}_1},m_{\chi^0_1})$ values
inside the region formed by joining the points (40,0), (60,30), (95,40),
(110,20), (85,0) and (40,0) GeV.  On the other hand, this region is excluded
in a model independent fashion once $m_{\tilde{t}_1}$ and $m_{\chi^0_1}$
are specified.
Here, we propose a method to improve the stop mass limit or discover
the stop squark by studying the decays of the top quark to a stop squark and
the lightest neutralino.

Alternatively, a light stop discovery or exclusion can
result from analyzing
the decay $t\rightarrow \tilde{t}_1 \chi^0_1\rightarrow c\chi^0_1\chi^0_1$.
We propose a search for this decay in $t\overline{t}$ production with
the associated standard decay $\overline{t}\rightarrow\overline{b}W^-$
(and the charge conjugate final state).  This final state provides the
distinctive signature $\ell^\pm~b~j~{\hbox to -2pt{/\hss}}E_T$, where
$\ell = e~\mbox{or}~\mu$ and $j$ is a non--$b$--jet.
Demanding a high--$p_T$, isolated lepton in the
central rapidity region of the detector guarantees a high triggering
efficiency and reduces backgrounds.  Also, since we demand a leptonic
decay of the $W$ boson
and because of the $t$ cascade decay to a charm jet and neutralinos,
the ${\hbox to -2pt{/\hss}}E_T$ spectrum is harder than for direct
stop pair production.

The outline of this paper is as follows:
In Sec. 2, we show how the CDF and {D\O} top quark data
set an upper bound on any non--SM decay mode (any mode other than
$t \rightarrow b W$)
of the top quark. Hence, BR($t \rightarrow {\tilde{t}_1} {\chi^0_1}$)
is bounded from above. In Sec. 3,
we explain the details of defining the signal and reducing potential
backgrounds for observing stop by studying $t\overline{t}$ events.
Discussion and conclusions are presented in Sec. 4.

\section{Upper Bound on The Branching Ratio of Non--SM Decay Modes of The
Top Quark}
\indent

The existence of the top quark is now firmly established \cite{CDF}.
Kinematic reconstruction of the decay products of the top quark in the decay
$t\rightarrow bW^+\rightarrow bjj$ (and the charge conjugate decay)
suggests a top mass
$m_t = 176\pm8\pm10$ GeV from the CDF data and
$m_t = 199^{+19}_{-21}\pm22$ GeV from
the D\O~data.
Both experiments have reported production cross sections,
which are a function of the assumed top mass used in the analysis.
It is important to remember that these experiments have optimized their
search for the process
$\mbox{p}\overline{\mbox{p}}\rightarrow t\overline{t}X\rightarrow b{W}^+
\overline{b}{W}^-X$, so they actually report the product of
the top production cross section $\sigma_{t\overline{t}}$ and the
branching ratio squared $b^2$, where $b$ = BR($t\rightarrow bW$).
Based on single-- and double--$b$--tagged events, CDF has
also reported a measurement of $b$~\cite{incandela}.
Finally, progress has been made
in understanding
the SM prediction for the production
cross section \cite{berger},
in which the
effects of multiple soft-gluon emissions have been properly resummed.

Since the measurement of the cross section obtained from the
``counting'' experiments (counting the observed
total $t\overline{t}$ event numbers in various decay modes)
and the measurement of the mass of the top quark (obtained from
reconstructing the invariant mass of the top quark) are not
strongly correlated,
one can combine these results to find the best fitted
values for $m_t$ and $\sigma_{t\overline{t}}$ \cite{soper}.
Using these results, we construct a $\chi^2$ function for
$m_* = m_t-176\mbox{~GeV}, \sigma_{t\overline{t}},$ and $b^2$:

\begin{eqnarray*}
{\chi}^2 =
\left(\frac{m_*}{12.8}\right)^2 +
\left(\frac{m_*-23}{29.7}\right)^2 +
\left(\frac{\ln(\sigma_{t\overline{t}}\times b^2/\sigma_{\rm CDF})}
    {\delta\ln\sigma_{\rm CDF}}\right)^2 + \\
\left(\frac{\ln(\sigma_{t\overline{t}}\times b^2/\sigma_{\rm D\O})}
    {\delta\ln\sigma_{\rm D\O}}\right)^2 +
\left(\frac{b^2-b^2_{\rm CDF}}{\delta b^2_{\rm CDF}}\right)^2 +
\left(\frac{\ln(\sigma_{t\overline{t}}/\sigma_{\rm th})}
    {\delta\ln\sigma_{\rm th}}\right)^2.
\end{eqnarray*}
In the above equation, we have taken the quadratic sum of the
statistical and systematic errors for measuring $m_t$.
$\sigma_{\rm CDF}$ and $\sigma_{\rm D\O}$ are
functional fits to the observed CDF and
D\O~$\sigma_{t\overline{t}}\times b^2$,
$\sigma_{\rm th}$ is a functional fit to the theoretical production
cross section, $b^2_{\rm CDF}$ is the measured BR($t\rightarrow bW$),
and all $\delta$'s are errors on these quantities, as listed below:

\begin{eqnarray*}
\begin{array}{lcl}
\ln\sigma_{\rm CDF}& = & \ln(7.60) - 3.17\times10^{-3}m_*
   -3.23\times10^{-5}m_*^2 - 2.94\times10^{-6}m_*^3 \\
\delta \ln\sigma_{\rm CDF}& = & \left\{{\begin{array}{cc} \frac{2.0}{7.6}
   & \sigma_{t\overline{t}} \le 7.6 {\rm pb} \\ \frac{2.4}{7.6} &
   \sigma_{t\overline{t}} > 7.6 {\rm pb}
                             \end{array} }\right. \\
\ln\sigma_{\rm D\O}& = & \ln(8.52) - 1.45\times10^{-2}m_*
   +8.87\times 10^{-5}m_*^2 \\
\delta \ln\sigma_{\rm D\O}& = & \frac{2.2}{6.4} \\
\ln\sigma_{\rm th}& = & \ln(5.38)-3.20\times10^{-2}m_*
   +3.65\times10^{-5}m_*^2 \\
\delta\ln\sigma_{\rm th}& = & .1 \\
b_{\rm CDF}& = & .87 \\
\delta b_{\rm CDF}& = & \left\{{\begin{array}{cc} .32 & b\le.87 \\
   .18 & b>.87 \end{array}}\right. .
\end{array}
\end{eqnarray*}

Finding the minimum value $\chi^2_{min}$ yields
$m_t = 168.6^{+3.0}_{-3.0}$ GeV, $\sigma_{t\overline{t}}
= 7.09^{+.68}_{-.62}$\,pb and $b = 1.00^{+.00}_{-.13}$.\footnote{
The theoretical prediction of $\sigma_{t\overline{t}}$ is
$6.83$\,pb for $m_t=168.6$\,GeV.}
At the 95\% confidence level (C.L.)
\footnote{We varied the parameter until the
$\chi^2$ value increased from $\chi^2_{min}$ by $(1.96)^2$ units.},
$b=.74$.  This
number, then, gives us an upper limit on BR($t\rightarrow X$), where
$X \ne bW$.
{}From the results of the fit described above, we
conclude that BR($t \rightarrow X$) for $X \neq bW$
has to be less than $\sim 25$\%.

Some comments are in order.  We have assumed that the experiments
observe $t\rightarrow bW$, where a fraction of the $b$ quarks have
been tagged.  However, there might be non--SM
decays involving $b$--quarks, such as $t\rightarrow bH^+$, where
$H^+$ is a charged Higgs boson.
Unfortunately, there
is at present no way to exclude such a scenario for the data used in
the fit described above when the mass of $H^+$ is about equal to
$M_W$.  The counting
experiment measurement of $\sigma_{t\overline{t}}$ is sensitive
to all events with a $b$--jet plus additional jets in the final state that
satisfy a certain set of cuts.
The top mass measurement only uses those events from the counting
experiment satisfying a $M_W$ mass constraint.  However, some of
the events in this mass distribution are background
events, which are subtracted statistically using the background estimated
by the counting experiment.  Therefore, we cannot define a subset of events
that is background free and excludes non--SM decays of the
top quark containing a $b$--quark.
On the other hand,
the mass distribution of the
non--$b$--jets for the counting experiment is consistent with
that expected from a $W$.  We continue our current analysis using
the assumption that the experiments do only measure
the $bW$ final state.
With a larger $t\overline{t}$ data sample in the near future,
it would be better to use double-$b$-tagged events
(from both $\ell+\,{\rm jets}$ and dilepton samples)
for the fit.

In the following section, we study how to directly observe the non--SM
decay mode of the top quark
$t \rightarrow {\tilde{t}_1} {\chi^0_1} \rightarrow
c {\chi^0_1}{\chi^0_1} $ and determine the minimum branching
ratio of this mode to be detected at the Tevatron for a given set of
the sparticle masses $m_{\tilde{t}_1}$ and $m_{\chi^0_1}$.

\section{Detecting Stop Squarks in the Decay of
Top Quarks in $t\overline{t}$ Pairs}
\indent

In this section, we consider the MSSM models in which
$m_t > m_{\tilde{t}_1}+ m_{\chi^0_1}$ and
$m_{\chi^\pm_1}+m_b$,
$M_W+m_{\chi^0_1}+m_b > m_{\tilde{t}_1} >m_{\chi^0_1}+m_c$, and
all other superpartners are heavier than $m_t$.
Hence, the dominant decay of $\tilde{t}_1$ is
$\tilde{t}_1\rightarrow c \chi^0_1$~\cite{bdggt}.
Since the branching ratios of the new decay modes are small ($<25\%$)
compared to the $bW$ final state, most top quarks
decay in the standard fashion (see Sec. 2).  As a result, we can
use the $b$--quark and a high--$p_T$ lepton to $tag$
$t\overline{t}$ production.  The non--SM decay of the top quark
$t \rightarrow {\tilde{t}_1} {\chi^0_1} \rightarrow
c {\chi^0_1}{\chi^0_1} $
provides an additional jet and missing transverse energy.
The signature of the $t\overline{t}$ events of interest
is thus $W(\rightarrow\ell\nu_\ell)+b+j+{\hbox to -2pt{/\hss}}E_T$.

Throughout this study, we assume the top mass is 175 GeV.
The top quark pair production cross section is given by the
QCD calculation, $\sigma_{t\overline{t}}(m_t=175\mbox{~GeV}) = 5.52$ pb
\cite{berger}.
For each model studied, we determine
the detection efficiencies  for the signal
$t(\rightarrow c \chi^0_1 \chi^0_1)
  \overline{t}(\rightarrow \overline{b} W^-(\rightarrow \ell^- \nu_\ell))$
(and the charge conjugate final state)
and the backgrounds.
The intrinsic backgrounds are $t(\rightarrow b W^+(\rightarrow X))
  \overline{t}(\rightarrow \overline{b} W^-(\rightarrow \ell^- \nu_\ell))$,
$ W^-(\rightarrow \ell^-\nu_\ell)X,$ (and the charge conjugate final
states) and
$Z(\rightarrow\ell^+\ell^-)X$.
For the signal rate, we include $\ell=e~\mbox{and}~\mu$. For
the background processes, we include $\ell=e,\mu,~\mbox{and}~\tau$
to account for the possible
large {\hbox to -2pt{/\hss}}$E_T$ background from $\tau$ decay.
We note that {\hbox to -2pt{/\hss}}$E_T$ can be faked if any
additional jet or lepton escapes detection.
The expected background event rate from $W$+jets, $Z$+jets, etc. is
calculated in the SM using {\tt PYTHIA 5.7}~\cite{pythia}.
The $W/Z$ + jets backgrounds are estimated from the $W/Z$ + parton
processes with a minimum $p_T$ = 20 GeV.
(The other backgrounds coming from a jet faking
an isolated lepton with high $p_T$
can be ignored after demanding also
a large ${\hbox to -2pt{/\hss}}E_T$ in the event.)
Signal events were generated for a set of
$(m_{\tilde{t}_1},m_{\chi^0_1})$ points
(see Table 1) beyond the D\O~search limit using an extended
version of {\tt PYTHIA 5.7}~\cite{spythia}.

Because of the presence of two neutralinos in the final state, we
expect that the signal will have a harder
${\hbox to -2pt{/\hss}}E_T$ spectrum
than the backgrounds.  Also, the correlation between the
${\hbox to -2pt{/\hss}}E_T$ and the lepton momentum in pure $W$ decays,
as observed in the transverse mass $m_T$ distribution,
should not be present in the signal.  The transverse mass $m_T$ is defined
by the expression $m_T^2 = 2 p_T^{(\ell)}{\hbox to -2pt{/\hss}}E_T
(1-\cos\Delta\phi_{\ell\nu})$, where $\Delta\phi_{\ell\nu}$ is the
azimuthal angle between the lepton $\ell$ and the
${\hbox to -2pt{/\hss}}E_T$ direction.
Finally, the hadronic activity should
be lower for the signal than for the $t\overline{t}$ background.
We find that the following cuts enhances the signal with respect to
the backgrounds:
\begin{itemize}
\item $p_T^{(\ell)} > 20~\mbox{GeV}, |\eta^{(\ell)}| < 1$. (I)
\item ${\hbox to -2pt{/\hss}}E_T > 60~\mbox{GeV}.$ (II)
\item $p_T^{(j)} > 15~\mbox{GeV}, |\eta^{(j)}| < 2$. (I)
\item $n_{jets} = 2$. Jets are defined by summing the transverse energy
$E_T$ in a toy calorimeter within a cone size $R=.7$ so that the jet
transverse energy $E^j_T > 15$ GeV. (III)
\item $m_T > 110~\mbox{GeV}.$ (II)
\item Fake~${\hbox to -2pt{/\hss}}E_T$ discrimination:
$\sqrt{(\pi-\Delta\phi_{j_1\nu})^2+(\Delta\phi_{j_2\nu})^2} > .5$,
where $j_1~\mbox{and}~j_2$ are the highest and second highest $E_T$ jets
and $\nu$ is the ${\hbox to -2pt{/\hss}}E_T$ direction, and
$\Delta\phi_{j\nu} > .1$ for all $j$. $\Delta\phi_{j\nu}$ is
the azimuthal angle between the jet $j$ and
the ${\hbox to -2pt{/\hss}}E_T$ direction. (I)
\end{itemize}
The cuts are separated into 3 sets (I--III) to show the behavior of
the signal and background.  Set (I) includes the minimal cuts
${\hbox to -2pt{/\hss}}E_T > 20~\mbox{GeV}$ and $n_{jets} \ge 2$.
The effect of these cuts are illustrated in Table 1 for 11 signal
$(m_{\tilde{t}_1},m_{\chi^0_1})$ points and the 3 major backgrounds.
For clarity, we have broken down the $t\overline{t}$ background into
three explicit final states:
{$b\overline{b}(e^\pm,\mu^\pm)\nu jj$},
{$b\overline{b}\tau^\pm\nu jj$}, and
{$b\overline{b}\ell^\pm\nu\ell^\mp\nu$}, where
$\ell = e,\mu,~\mbox{and}~\tau$.
We find that the major background comes from the dilepton
decays, {$b\overline{b}\ell^\pm\nu\ell^\mp\nu$},
where one lepton escapes detection.
Note that the signal
detection efficiencies $\epsilon_{\rm S}$ are about the same,
$\sim .06-.08$ for various choices
of $(m_{\tilde{t}_1},m_{\chi^0_1})$  points.
Also in Table 1, we present similar results using a wider
pseudorapidity range: $|\eta^{(\ell)}|, |\eta^{(j)}| < 2.5$.
We find that the detection efficiencies for the backgrounds do not change
significantly, while the signal efficiencies increase by about 30\%.
\begin{table}
\renewcommand{\arraystretch}{1.33}
\begin{center}
\begin{tabular}[bht]{|c|c||c|c|c||c|c|c|}\hline
$m_{\tilde{t}_1}$ & $m_{\chi^0_1}$ &
\multicolumn{3}{c||}{\bf Efficiency after cuts} &
\multicolumn{3}{|c|}{\bf Efficiency after cuts}\\
                  &                &
\multicolumn{3}{c||}{                         } &
\multicolumn{3}{|c|}{\bf With Larger $\eta$ Acceptance}\\
\cline{3-8}
(GeV) & (GeV) & I & II & III & I & II & III \\ \hline
50 & 20 & .229 & .094 & .068  & .353 & .140 & .103 \\
50 & 30 & .208 & .088 & .063  & .332 & .131 & .097 \\
50 & 40 & .154 & .073 & .056  & .259 & .104 & .075 \\
60 & 30 & .226 & .095 & .071  & .319 & .127 & .092 \\
60 & 40 & .201 & .086 & .065  & .286 & .116 & .085 \\
80 & 40 & .235 & .100 & .075  & .286 & .110 & .078 \\
80 & 60 & .186 & .081 & .060  & .218 & .097 & .072 \\
90 & 50 & .243 & .102 & .076  & .342 & .132 & .095 \\
90 & 70 & .174 & .082 & .062  & .250 & .107 & .078 \\
100& 40 & .248 & .102 & .076  & .357 & .137 & .101 \\
100& 60 & .239 & .100 & .076  & .332 & .129 & .095 \\
\hline
\multicolumn{2}{|l||}{$t\overline{t}~~~~~\sigma=5.52~\mbox{pb}$ }
& .150 & .011 & .006 & .193 & .013 & .006
\\ \cline{3-8}
\multicolumn{2}{|c||}{$b\overline{b}(e^\pm,\mu^\pm)\nu jj$ }
& .375 & .004 & .0   & .506 & .005 & .0
\\
\multicolumn{2}{|c||}{$b\overline{b}\tau^\pm\nu jj  $ }
& .001 & .0 & .0 & .001 & .0 & .0 \\
\multicolumn{2}{|c||}{$b\overline{b}\ell^\pm\nu\ell^\mp\nu   $ }
& .347 & .095 & .005 & .388 & .107 & .005 \\ \hline
\multicolumn{2}{|l||}{$ZX~~\sigma=105~\mbox{pb}$}
& .002 & $<2\times10^{-5}$ & $<2\times10^{-5}$
& .004 & $<2\times10^{-5}$ & $<2\times10^{-5}$ \\
\multicolumn{2}{|l||}{$WX~\sigma=775~\mbox{pb}$}
& .043 & $1\times10^{-5}$ & $1\times10^{-5}$
& .081 & $1\times10^{-5}$ & $1\times10^{-5}$ \\
\hline
\end{tabular}
\end{center}
\caption{Efficiency of Cuts (I--III) for Signal and Background.}
\end{table}

Given the efficiency for detecting the signal
$\epsilon_{\rm S}$ and the backgrounds $\epsilon_{\rm B}$, their
respective rates are a function of the branching ratio
BR($t\rightarrow c \chi^0_1 \chi^0_1)\equiv b_X$.
For the models considered,
there are no other ``new'' decay modes of
$t$, thus BR($t\rightarrow bW)= 1 - b_X$.
The signal rate is
$\sigma_{t\overline{t}}\times 2\times b_X\times(1-b_X)\times2/9
\times\epsilon_{\rm S}$,
where the factor of 2/9 accounts for the $e$ and $\mu$
decay modes of the $W$.
The background rate from $t\overline{t}$ is
$\sigma_{t\overline{t}} \times(1-b_X)^2\times
\epsilon_{t\overline{t}}$.
The other background rates are simply a product of the production
cross section and their efficiencies.\footnote{As shown in Table 1,
after the cuts I--III, the backgrounds $Z(\rightarrow\ell^+\ell^-)X$ and
$W^\pm(\rightarrow \ell^\pm \nu_\ell)X$, where $\ell=e,\mu,~\mbox{and}~\tau$,
are  2.1\,fb and 7.8\,fb, respectively.}
Comparing the data with these predictions, one can then
set an upper bound on $b_X$
for any given ($m_{\tilde{t}_1}$, $m_{\chi^0_1}$), and,
therefore, can constrain
the predicted allowed models from Ref.~\cite{stuart} if no
signal is found.

{}From the discussion in Sec. 2, we conclude that $b_X < 25\%$ at the
95\% C.L. inferred from a global fit to the CDF and D\O~data (assuming
the entire data sample to be $t\rightarrow bW$ events).
Here,
we would like to know the minimum $b_X$ that can be directly measured
(in contrast to that inferred from fitting) by
detecting the $t\overline{t}$ pair events at the Tevatron
for a given integrated luminosity (${\cal L}$)
of the collider, if we demand a 3--$\sigma$ effect.
For simplicity
(later, we perform a more thorough analysis), let
us consider only the major background
from the SM decays of $t\overline{t}$ pairs.
Define the number of signal $N_S$ and background $N_B$ events:
\begin{eqnarray}
N_S & = & {\cal L} \times \sigma_{t\overline{t}} \times 2
\times b_X \times (1-b_X) \times 2/9\times \epsilon_{\rm S} \, ,
\nonumber \\
N_B & = & {\cal L} \times \sigma_{t\overline{t}} \times (1-b_X)^2
\times \epsilon_{t\overline{t}}.
\end{eqnarray}
Assuming that Gaussian statistics are applicable,
a 3--$\sigma$ effect of the signal over background requires
\begin{equation}
{N_S \over \sqrt{N_B} }= 4/9 \times b_X \times
\sqrt{{\cal L} \cdot \sigma_{t\overline{t}}}
{ \epsilon_{\rm S} \over \sqrt{\epsilon_{t\overline{t}}} } \ge 3 \, .
\nonumber
\end{equation}
Taking $\sigma_{t\overline{t}}=5.5$\,pb, ${\cal L}=2\,{\rm fb}^{-1}$,
$\epsilon_{\rm S}=0.1$ and $\epsilon_{t\overline{t}}=0.01$,
the above equation gives $b_X \ge 6\%$.
Substituting $b_X$ into Eq. 1, one would expect to observe approximately
$N_S=30$ and $N_B=100$ events in the 2\,${\rm fb}^{-1}$ data sample
if BR($t\rightarrow c\chi^0_1\chi^0_1$)~=~.06.
Although the ratio of signal to background is about 1 to 3,
the distributions in $m_T$ for the signal and the background events
are very different.  A few examples of $m_T$ distributions for the signal
and background are given in Fig.~1.
The distributions shown have passed all cuts
I--III except for the $m_T$ cut, and have been normalized to have the
same unit area.  The $t\overline{t}$ background is denoted by large
hatches, $W$+jets with small hatches, and three representative signal points
($m_{\tilde{t}_1}$, $m_{\chi^0_1}$)
have clear regions outlined by solid ((50,30) GeV),
dot--dashed ((90,50) GeV) and dashed ((100,60) GeV) lines.
$m_T$ is larger for a heavier ${\chi^0_1}$
and, in that case, the background event can be easily
distinguished from the signal event.

\begin{figure}[hbt]
\centering \psfig{file=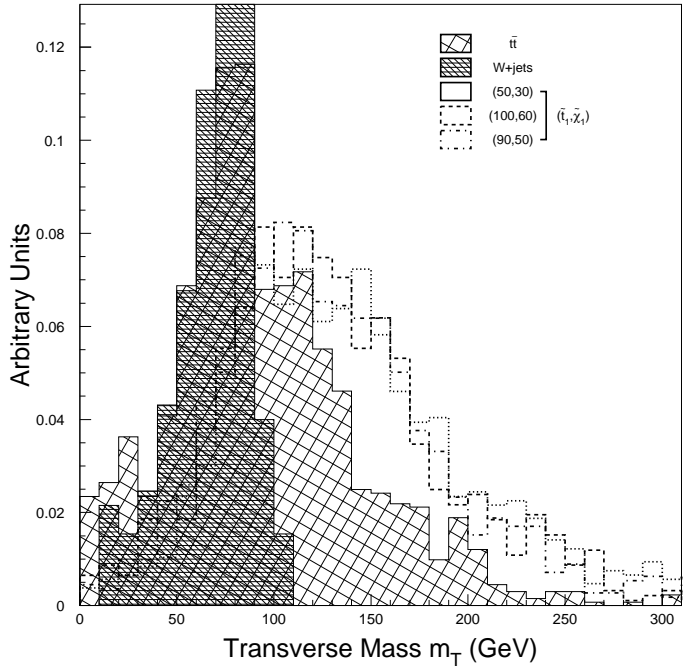,width=10cm}
\caption{Transverse Mass $m_T$ Distribution for Signals and Backgrounds}
\end{figure}

Since the detection efficiency of the signal is not sensitive to the
masses of ${\tilde{t}_1}$ and ${\chi^0_1}$, the limit on $b_X$ can either
confirm the MSSM predictions by finding a stop squark or
exclude the corresponding parameters of the model.
The branching ratio that can be probed for 200 pb$^{-1}$, 2 fb$^{-1}$,
and 10 fb$^{-1}$ is
listed in Table 2 including all backgrounds and using Poisson statistics
where applicable.
To summarize Table 2, the worst $b_X$ reach for the models
considered is .45, .10, and .04 for 200 pb$^{-1}$, 2 fb$^{-1}$ and
10 fb$^{-1}$.  With the increased pseudorapidity coverage, the reach is
extended to .33, .08, and .04 for 200 pb$^{-1}$, 2 fb$^{-1}$ and
10 fb$^{-1}$.
The value of $b_X$ reachable in the present data sample is
comparable to the indirect limit from the cross section and
mass measurements (see Sec. 2),\footnote{
The 3--$\sigma$ bound on $b$ from the global fit discussed in Sec. 2 is
.61.  Consequently, the bound on $b_X$ is .39.}
and clearly consistent with the
measurement of the branching ratio from CDF.

\begin{table}
\renewcommand{\arraystretch}{1.33}
\begin{center}
\begin{tabular}[hbt]{|c|c||c|c|c||c|c|c|}\hline
$m_{\tilde{t}_1}$ & $m_{\chi^0_1}$ &
\multicolumn{3}{c||}{\bf $b_X$ vs. Luminosity }  &
\multicolumn{3}{|c|}{\bf $b_X$ vs. Luminosity } \\
      &       &
\multicolumn{3}{c||}{  } &
\multicolumn{3}{|c|}{\bf With Larger $\eta$ Acceptance }\\ \cline{3-8}
(GeV) & (GeV) & 200 pb$^{-1}$ & 2 fb$^{-1}$ & 10 fb$^{-1}$  &
200 pb$^{-1}$ & 2 fb$^{-1}$ & 10 fb$^{-1}$ \\ \hline
50 & 20 & .33  & .08  &  .04 & .22 & .06 & .03 \\
50 & 30 & .35  & .09  &  .04 & .22 & .06 & .03 \\
50 & 40 & .45  & .10  &  .04 & .32 & .07 & .03 \\
60 & 30 & .33  & .08  &  .04 & .23 & .06 & .03 \\
60 & 40 & .33  & .08  &  .04 & .26 & .06 & .03 \\
80 & 40 & .32  & .07  &  .03 & .30 & .07 & .03 \\
80 & 60 & .39  & .09  &  .04 & .33 & .08 & .04 \\
90 & 50 & .31  & .07  &  .03 & .22 & .06 & .03 \\
90 & 70 & .36  & .09  &  .04 & .30 & .07 & .03 \\
100& 40 & .31  & .07  &  .03 & .22 & .06 & .03 \\
100& 60 & .31  & .07  &  .03 & .22 & .06 & .03 \\
\hline
\end{tabular}
\end{center}
\caption{Limit on Branching Ratio $b_X$ as a Function of Luminosity for
the Models Studied.}
\end{table}

\section{Discussion and Conclusions}
\indent

{}From a global fit to the available data for the
top quark mass, production cross section, and SM branching ratio and the
predicted top production cross section, we have determined the allowed
non--SM branching ratio for top quark decay.  This branching ratio is
bounded to be less than about 25\% at the 95\% C.L.
Since this branching ratio is
small, we studied the possibility of observing the rare decay
of
$t\rightarrow \tilde{t}_1 \chi^0_1$
in association with the SM decay
$\overline{t}\rightarrow b W^-
\rightarrow b\ell^-\nu$ (and the charge conjugate decays).
Additionally, we required
$m_{\chi^\pm_1}+m_b,M_W+m_{\chi^\pm_1}+m_b >
m_{\tilde{t}_1} > m_{\chi^0_1}+m_c$,
so that the dominant
stop squark decay mode is $\tilde{t}_1\rightarrow c\chi^0_1$.
For the models studied, we found that the signal detection
efficiency is approximately constant and almost
independent of the stop or neutralino mass for
the mass region considered.
If no signal is found,  we could exclude
models with BR($t \rightarrow \tilde{t}_1 \chi^0_1$) larger
than .33, .08 and .04 at the 3--$\sigma$ level
for the Tevatron with a luminosity of
200\,${\rm pb}^{-1}$, 2\,${\rm fb}^{-1}$ and 10\,${\rm fb}^{-1}$,
respectively.
If a signal is found, the signal (background) event yields for the smallest
$b_X$ are 8(5), 26(80), and 68(430) for a 200\,${\rm pb}^{-1}$,
2\,${\rm fb}^{-1}$ and 10\,${\rm fb}^{-1}$ data sample.

If, contrary to our assumptions about the lightest chargino,
$m_{\chi^\pm_1}+m_b < m_{\tilde{t}_1}$, then the
two body decay $\tilde{t}_1\rightarrow b\chi^+_1$ dominates.
For models discussed in Ref.~\cite{stuart},
a light chargino would preferentially decay via
 $\chi^\pm_1\rightarrow f\overline{f^{'}}\chi^0_1$.
The final state is similar to that from the SM decay, but the $\chi^\pm_1$
decay products are softer and have a lower acceptance.
It is therefore better to detect the light chargino directly from
chargino pair production than from the top quark decay.

In conclusion, we find that the prospect to observe a light stop at
the Tevatron is very promising.
Unless a stop signal can be found in the current or upcoming
data at the Tevatron, it will be difficult to reconcile experiment with
low energy Supersymmetry in the case that
$t \rightarrow {\tilde{t}_1} {\chi^0_1} \rightarrow
c {\chi^0_1}{\chi^0_1} $
is the dominant non--SM decay mode.

\section*{ Acknowledgments }
The authors thank E.L.~Berger,
G.L.~Kane, J.~Linnemann, L.~Nodulman, J.D.~Wells, and A.B.~Wicklund
for useful discussions.
S.M. was supported in part by DOE grant DE--FG03--92--ER40701.
The work of C.P.Y. was supported in part by NSF grant No. PHY-9309902.

\end{document}